\documentclass[A4, 11pt]{article}
\usepackage{fullpage}
\usepackage{caption}
\usepackage{subcaption}

\usepackage[authoryear,round]{natbib}
\usepackage{graphicx}

\usepackage{bm}
\usepackage[parfill]{parskip}

\usepackage[authoryear,round]{natbib}
\usepackage{amsmath}
\usepackage{amsthm}

\usepackage{hyperref}

\newcommand{\x}{\bm{x}}
\newcommand{\bphi}{\bm{\phi}}
\newcommand{\bmu}{\bm{\mu}}
\newcommand{\y}{\bm{y}}

\graphicspath{{figures/}}

\title{Understanding the 2016 US Presidential Election using ecological inference and distribution regression with census microdata}
\author{Seth Flaxman (Department of Statistics, Oxford)\footnote{For correspondence: flaxman@stats.ox.ac.uk},\\
Danica J.\ Sutherland (Gatsby Computational Neuroscience Unit, University College London), \\
Yu-Xiang Wang (Machine Learning Department, Carnegie Mellon University), \\
and Yee Whye Teh (Department of Statistics, Oxford)}

\begin{document}
\maketitle
\abstract{We combine fine-grained spatially referenced census data with the vote outcomes from the 2016 US presidential election. Using this dataset, we perform ecological inference using distribution regression (Flaxman et al, KDD 2015) with a multinomial-logit regression so as to model the vote outcome Trump, Clinton, Other / Didn't vote as a function of demographic and socioeconomic features. Ecological inference allows us to estimate ``exit poll'' style results like what was Trump's support among white women, but for entirely novel categories. We also perform exploratory data analysis to understand which census variables are predictive of voting for Trump, voting for Clinton, or not voting for either. All of our methods are implemented in Python and R, and are available online for replication.}

\section{Introduction}

The results of the 2016 US Presidential Election were, to put it mildly, a surprise.
Pre-election polls and forecasts based on these polls pointed to a Clinton victory, a
prediction shared by betting markets and pundits. In the aftermath of the vote, the main 
question asked is ``why?'' with answers ranging from the political to the economic to the
social/cultural.  In this article we attempt to provide a preliminary answer to a fundamental
question: who voted for Trump, who voted for Clinton, and who voted for a third party or did not
vote? 

By combining data from the United States census with the election results and
recently proposed machine learning methods for ecological inference using
regressions based on samples from a distribution, we provide local demographic
estimates of voting and non-voting.  Unlike with exit polls, we are able to
draw conclusions across the entire US and at a local level, about voters and
non-voters, for interesting and novel combinations of predictor variables.  It
is our hope that this analysis will help inform the typical election post
mortems, which are usually informed by incomplete information, due the
following factors:
\begin{itemize}

	\item Vote counts will not be finalized in many precincts until days or
		in rare cases weeks after the election. Very close popular vote
		totals yield winner-take-all results, a fact of the US's
		electoral system but one that can lead to winner-take-all
		explanations.

	\item The winner-take-all system means that the geography of voting
		patterns is often overly simplified, with states considered as
		homogeneous entities, leading to the ubiquitous red and blue
		maps popularized in 2000 \citep{gastner2005maps}. The
		geography of voting patterns are also often based on overly
		broad demographic/geographic categories, such as ``suburban
		whites'' which do not necessarily reflect an up-to-date picture of
		who lives where.

	\item Exit polls (surveys of voters) conducted by Edison Research on
		behalf of a consortium of media outlets are available
		immediately, with sufficient sample coverage to yield
		representative samples in 28 US states and nationally. But they
		are not designed to provide sub-state geographic coverage, and there
		is a limit to the number and types of questions they can ask, and
		questions about their reliability and coverage.
\end{itemize}
All of the source code for our
analysis is freely available online to enable
replication\footnote{\url{www.github.com/djsutherland/pummeler} and
\url{www.github.com/flaxter/us2016}}. 

Our analysis is in two parts: building on \cite{flaxman2015ecological}, we
provide exit poll style estimates of the number of voters and
non-voters by candidate and various subgroups. We are able to make much more
finely-grained estimates, at the local level, than conventional exit polling.
We make these estimates at the national, state, and local level. For example,
Table \ref{table:sex} shows that 47\% of voters were men while 53\% were women.
Among women, Clinton received 56\% of the vote while Trump received 45\% of the
vote. Unlike exit polls, we also estimate the participation rate by
demographic, calculating the fraction of ``potential voters''---voting-age
citizens---who voted for one of the two major party candidates. Equivalently,
we calculate the fraction of Other / non-voting. Thus we see that 50\% of
eligible men participated, voting for either Clinton or Trump, while 53\% of
eligible women participated.

The second part of our analysis is an exploratory analysis of which group-level
variables (i.e.~distribution of age, joint distribution of income and race)
were correlated with the final outcome. While previous approaches to this question
have focused on correlations between group-level means\footnote{\url{http://www.nytimes.com/2016/03/13/upshot/the-geography-of-trumpism.html}},
our novel approach makes it possible to investigate correlations between the full 
distributions of these group-level variables, observed at the local area level.

Note that in the months after the election, high quality survey results will
become available (the 2016 American National Election Study will be a
large-scale look at who voted and why, and the US Census Bureau collects data
on who voted), along with voter rolls data, which will ultimately give a very
accurate portrait of who voted and who did not. In the meantime, we are left with
incomplete information, and it is our goal to begin to fill in part of the picture.

\section{Methods}
Given $n$ local areas of interest 
(counties reporting electoral results merged with PUMA regions, following
\cite{flaxman2015ecological}) we assume that we have:
\begin{itemize}
    \item Samples of individual level demographic data from the American Community Survey:
        \begin{equation}
            \{\x_1^j\}_{j=1}^{N_1}, \ldots, \{\x_n^j\}_{j=1}^{N_n}
        \end{equation}
%    \item Spatial coordinates $\s_1, \ldots, \s_n$, and
    \item Election results $\y_1, \ldots, \y_n$.
\end{itemize}
Election results for each local area $i$ are summarized as counts vectors $\y_i$.
We use the method of ``distribution regression'' \citep{szabo2015two,lopez2015towards,flaxman2015ecological}
to regress from a set of samples $\{\x_i^j\}_{j=1}^{N_i}$ to a label $\y_i$.
This entails encoding $N_i$ samples as a high-dimensional feature vector, using
explicit feature vectors and kernel mean embeddings \citep{smola2007hilbert,gretton2012kernel}:

We use an additive featurization to compute mean embeddings $\bmu_i$ as follows for local
area $i$:
\begin{align}
    \{\x_i^j\}_{j=1}^{N_i}  = \{\x_i^1, \x_i^2, \ldots, \x_i^{N_i} \}
\end{align}
Each $\x_i^j$ is a vector of length $d$ consisting of a mix of categorical and real
valued variables:
\begin{equation}
    \x_i^j = [x_{r1}^j, \ldots, x_{rd}^j]^{\top}
\end{equation}
We consider feature mappings $\phi_1, \ldots, \phi_d$ and the additive featurization
in which feature vectors are concatenated:
\begin{equation}
    \bphi(x_i^j) := [\phi_1(x_{r1}^j), \ldots, \phi_d(x_{rd}^j)]^{\top}
\end{equation}
We also consider including interaction terms of the form:
$\phi_{pq}(x_{rp}^j, x_{rq}^j)$ in $\bphi(x_i^j)$ in $\bphi$.

We use $\bphi$ to estimate mean embeddings, one for each region, where the
weights $w_j$ are the person weights, one per observation, reported in the census:
\begin{align}
    \bmu_i = \sum_{j=1}^{N_i} w_j \bphi(\x_i^j)
\end{align}

We model the outcome $\y_i$ as a function of covariate distribution vectors $\bmu_i$ using
penalized multinomial regression with a softmax link function. Let 
$$\y_i = [\mbox{Clinton votes}, \mbox{Trump votes}, \mbox{Non-votes and third party votes}]^{\top}$$ Then:
\begin{equation}
	\y \sim \mbox{Multinomial}(\mbox{softmax}(\bmu \beta_1, \bmu\beta_2, \bmu \beta_3))
\end{equation}
where $\mbox{softmax}$ generalizes the logistic link to multiple categories:
\begin{equation}
	\mbox{softmax}_i(x_1,x_2,x_3) = \frac{\exp(x_i)}{\exp(x_1) + \exp(x_2) + \exp(x_3)}
\end{equation}
We implemented the featurization using orthogonal random Fourier features \citep{felix2016orthogonal}
for real-valued variables and unary coding for categorical variables. We fit
the penalized multinomial model using {\tt glmnet} in R, crossvalidating to choose the $\alpha$
parameter (relative strength of $L_1$ vs.~$L_2$ penalty) and sparsity parameter $\lambda$.
We used {\tt glmnet}'s built-in group lasso functionality meaning that 
$(\beta_1)_i, (\beta_2)_i, (\beta_3)_i$ would either all be simultaneously
zeroed out by the $L_1$ penalty or not.

\subsection{Inferring who voted}
After obtaining maximum likelihood estimates $\hat \beta_1, \hat \beta_2, \hat \beta_3$
from our fit of the model, we used subgroup populations to make predictions, calculating
mean embeddings for each local area based only on the group under consideration.
For example, to predict the vote among women in region $i$ we calculated new predictor vectors
restricting our summation to the $N_i^{\mbox{w}}$ women in area $i$:
\begin{align}
	\bmu_i^{\mbox{w}} = \sum_{j=1}^{N_i^{\mbox{w}}} w_j \bphi(\x_i^j)
\end{align}
and then used $\hat \beta_1, \hat \beta_2, \hat \beta_3$ and $\mbox{softmax}$ to calculate
the corresponding probabilities of supporting Clinton, Trump, and other among women in 
region $i$. To calculate expected vote totals, we multiply these probabilities by
the estimated total number of women in each region i, calculated by summing the census weights:
\begin{equation}
\sum_j^{N_i^{\mbox{w}}} w_j
\end{equation}

\subsection{Group-based exploratory data analysis}
While exit poll style ecological inference can give us deep insights into various
preselected demographic categories, exploratory data analysis can reveal unexpected
patterns. We consider fitting the same models as above, but only using related
subsets of the features, e.g.~all of the categorical variables related to race
or the interaction between age and income. 

\section{Data and implementation}
Our analysis can be replicated using a Python package we wrote called {\tt
pummeler}\footnote{\url{www.github.com/djsutherland/pummeler}} with R
replication scripts in our GitHub
repository\footnote{\url{www.github.com/flaxter/us2016}}.  We obtained the most
recent American Community Survey 5-year dataset, 2010-2014 and the 2015 1-year
dataset and merged them.  We excluded data from 2010 and 2011 because these
years used the 2000 US Census geography, leaving us with four years of
individual-level data. We adjusted the 2015 weights to match 2012-14 thus
obtaining a 4\% sample of the US population, consisting of 9,222,637 million
individual observations.  Electoral results by county were scraped from nbc.com
on 9 November 2016 (the day after the election). Using the merging strategy
described in \cite{flaxman2015ecological} we ended up with 979 geographic
regions.  Real-valued data was standardized to have mean zero and variance one
and categorical variables were coded in unary, omitting a reference category to
aid in model fit.

We obtained state-level exit polling data obtained from foxnews.com for the following 
demographics: age, sex, and education. We used these to increase the size and diversity of our sample.
In addition to the 979 geographic regions labeled with election outcome data, we used
a total of 249 state-level subgroups, e.g.~women in Florida, calculating the subgroup feature
vectors as described above by restricting to the individual's in the census matching the
demographic reported in the exit poll. As shown in Figure \ref{fig:scatter} in the
Appendix, our model
is not overfitting, either to true outcome data (black) or exit poll data (blue). The 
best $\alpha$ parameter we found was 0.05 (where $0$ is pure ridge regression).
The number of parameters in the best model was 415 out of a total of 11,112 features.
Using this model, we made predictions for a variety of groups derived from the census,
as detailed in the next section.

\section{Results: inferring who voted with ecological inference}
\label{section:inferring}
\begin{figure}[ht]
    \centering
    \includegraphics[width=.9\textwidth]{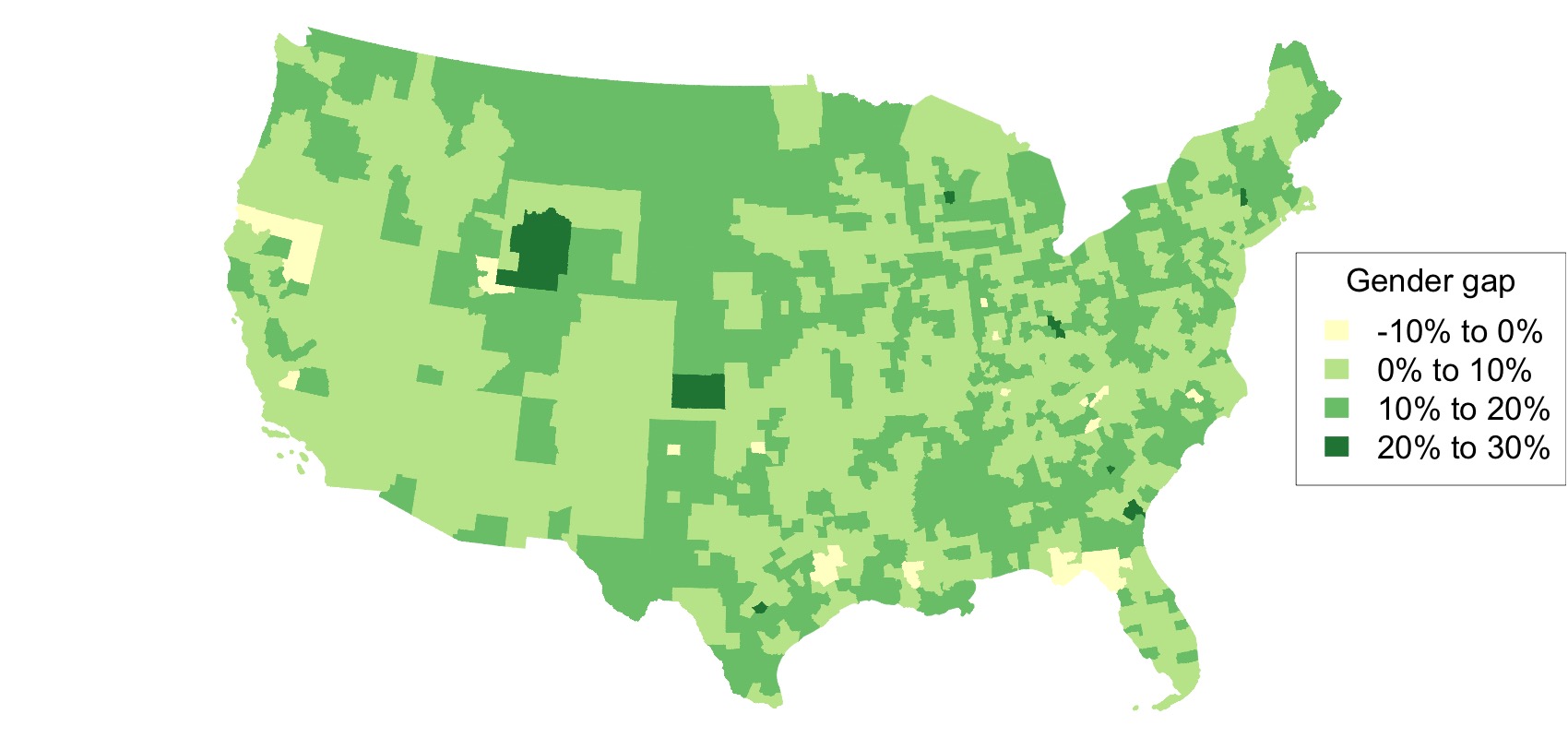}
    \caption{Gender gap in support for Trump represented as percentage point difference between
support for Trump among men minus support for Trump among women}
    \label{fig:gender-gap}
\end{figure}

We start by highlighting the local inferences that our method allows. We inferred the gender
gap of support for Trump among men minus support for Trump among women (11 percentage points
nationally) for each of 979 geographic regions. As shown in Figure \ref{fig:gender-gap}
gender gap varies spatially, with some regions showing parity and others exceeding the national
average---29 regions had gaps of more than 20 percentage points and 54 regions had negative gender gaps
(higher support for Clinton among men than women).

In this section we show exit poll-style results which we inferred using the
ecological inference methods described above. We could have presented these as
contingency tables, but we prefer to mimic the way that exit polls are often
presented in the popular press. Here is how to read Table \ref{table:sex}, for example.
Nationally, among voters who voted for either Trump or Clinton, we estimate that 45\% 
of men voted for Clinton while 55\% of men voted for Trump. These two columns will always
sum to 1 within a row. The third column says that we estimated that men made up 47\% of
the electorate, while women made up 53\% of the electorate. The column ``Participation Rate''
gives the percent of voting-age men/women who voted for either Clinton or Trump. The
column Other / non-voting is 1-Participation Rate as it gives the percentage of voting-age
men/women who voted for a third party or did not vote.

The tables give national results for categories usually reported in exit polls. 
A variety of other groupings are included in the Appendix in Section \ref{sec:supplement}
CSV files with all of our estimates are available at the local level on our github repository. 

% latex table generated in R 3.3.1 by xtable 1.8-2 package
% Thu Nov 10 00:42:23 2016
\begin{table}[ht]
\centering
\begin{tabular}{lrrrrr}
  \hline
 &  Clinton & Trump & Fraction of electorate & Participation Rate & Other / Non-voting \\ 
  \hline
Men  & 0.45 & 0.55 & 0.47 & 0.50 & 0.50 \\ 
  Women  & 0.56 & 0.44 & 0.53 & 0.53 & 0.47 \\ 
   \hline
  \hline
 18--29 year olds & 0.62 & 0.38 & 0.17 & 0.42 & 0.58 \\ 
   30--44 & 0.54 & 0.46 & 0.25 & 0.54 & 0.46 \\ 
   45--64 & 0.46 & 0.54 & 0.39 & 0.58 & 0.42 \\ 
   65  and older & 0.45 & 0.55 & 0.18 & 0.47 & 0.53 \\ 
   \hline
\end{tabular}
\caption{Estimated demographic trends using our model. Columns show support for Clinton, Trump, 
fraction of the electorate made up by men (first row) vs women (second row), 
fraction of voting-age citizens who voted for Clinton or Trump,
and fraction of voting-age citizens who voted for a third party or did not vote.}
\label{table:sex}
\end{table}

\begin{table}[ht]
\centering
\begin{tabular}{p{5cm}rrp{2cm}p{2cm}r}
  \hline
 &  Clinton & Trump & Fraction of electorate & Participation Rate & Other / Non-voting \\\hline
  Educational attainment & & & & & \\ 
  \hline
 High-school or less & 0.47 & 0.53 & 0.20 & 0.25 & 0.75 \\ 
   Some college & 0.46 & 0.54 & 0.33 & 0.49 & 0.51 \\ 
   Bachelor's degree & 0.51 & 0.49 & 0.30 & 0.84 & 0.16 \\ 
   Postgraduate education & 0.58 & 0.42 & 0.17 & 0.85 & 0.15 \\ 
   \hline
   \hline
  Race/Ethnicity & & & & & \\ 
  \hline
 hispanic & 0.68 & 0.32 & 0.09 & 0.43 & 0.57 \\ 
   white & 0.40 & 0.60 & 0.72 & 0.54 & 0.46 \\ 
   black & 0.84 & 0.16 & 0.14 & 0.58 & 0.42 \\ 
   asian & 0.76 & 0.24 & 0.03 & 0.39 & 0.61 \\ 
   \hline
   \hline
  Place of birth & & & & & \\ 
  \hline
 Born in the U.S. & 0.42 & 0.58 & 0.93 & 0.62 & 0.38 \\ 
   U.S. citizen by naturalization & 0.74 & 0.26 & 0.06 & 0.42 & 0.58 \\ 
   \hline
  \hline
  Personal income & & & & & \\ 
  \hline
 below \$50,000 & 0.53 & 0.47 & 0.64 & 0.44 & 0.56 \\ 
   \$50,000 to \$100,000 & 0.46 & 0.54 & 0.25 & 0.72 & 0.28 \\ 
   more than \$100,000 & 0.45 & 0.55 & 0.11 & 0.77 & 0.23 \\ 
   \hline
\end{tabular}
\caption{Estimated demographic trends using our model for various census variables. Columns show support for Clinton, Trump, 
fraction of the electorate made up by men (first row) vs women (second row), 
fraction of voting-age citizens who voted for Clinton or Trump,
and fraction of voting-age citizens who voted for a third party or did not vote.}
\label{table:other}
\end{table}

\clearpage

\section{Results: group-based exploratory data analysis}
\label{section:group-based}
\begin{figure}[p]
\begin{subfigure}{.45\textwidth}
    \centering
    \includegraphics[width=\textwidth]{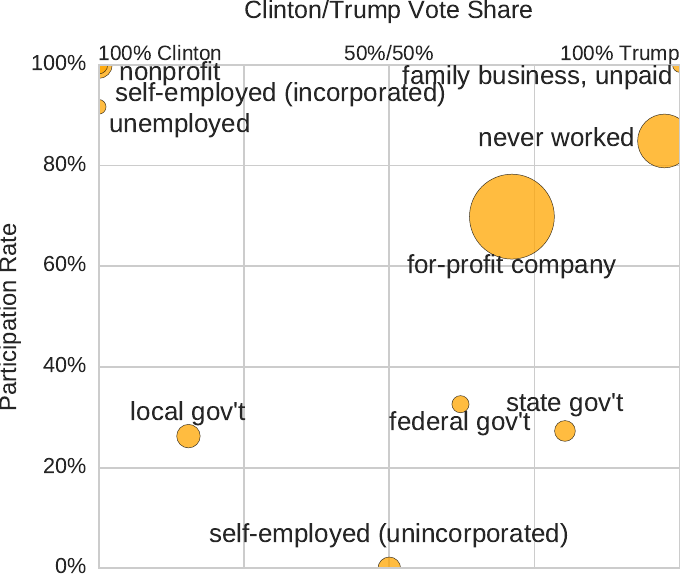}
    \caption{Class of worker {\tt COW}}
\end{subfigure}
~
\begin{subfigure}{.45\textwidth}
    \centering
    \includegraphics[width=\textwidth]{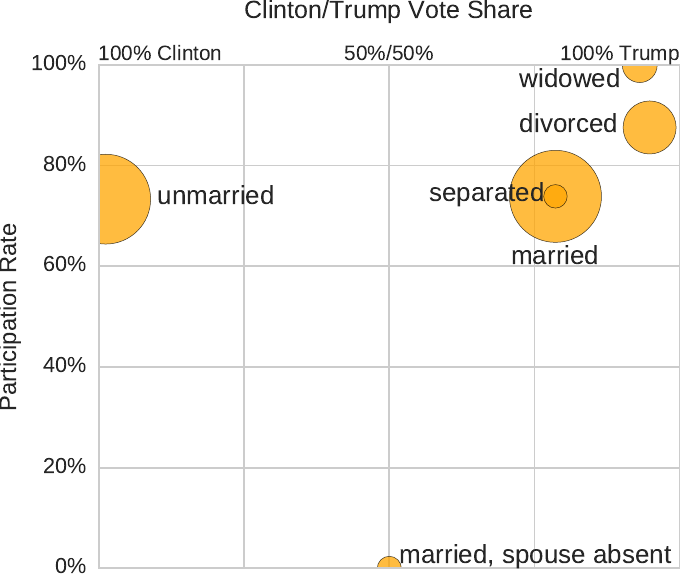}
    \caption{Marital status {\tt MSP}}
    \label{fig:marital}
\end{subfigure}
\\
\begin{subfigure}{.45\textwidth}
    \centering
    \includegraphics[width=\textwidth]{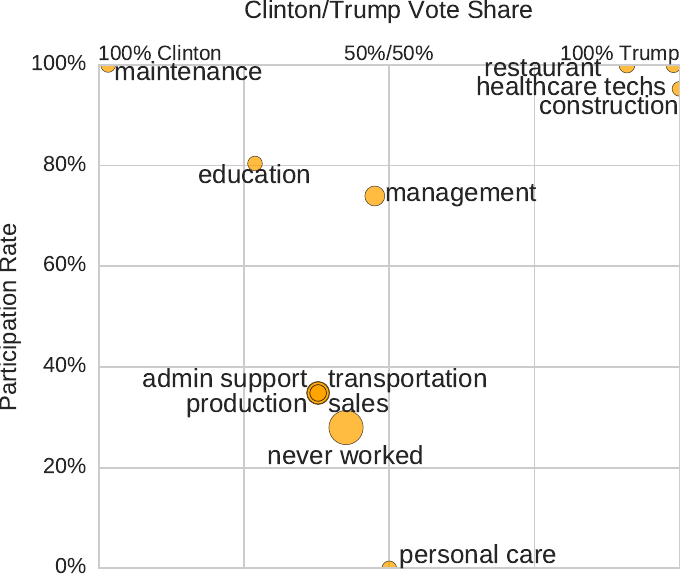}
    \caption{Occupation {\tt OCCP}}
\end{subfigure}
~
\begin{subfigure}{.45\textwidth}
    \centering
    \includegraphics[width=\textwidth]{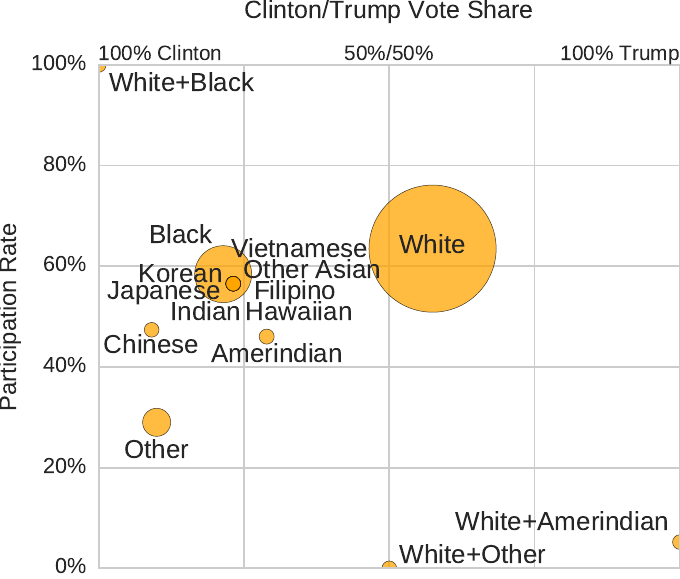}
    \caption{Race {\tt RAC3P}}
\end{subfigure}
\\
\begin{subfigure}{.45\textwidth}
    \centering
    \includegraphics[width=\textwidth]{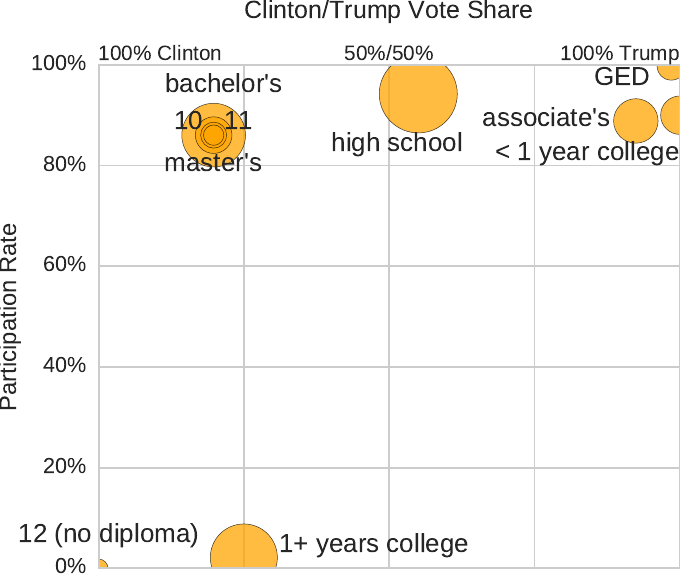}
    \caption{Education level {\tt SCHL}}
\end{subfigure}
~
\begin{subfigure}{.45\textwidth}
    \centering
    \includegraphics[width=\textwidth]{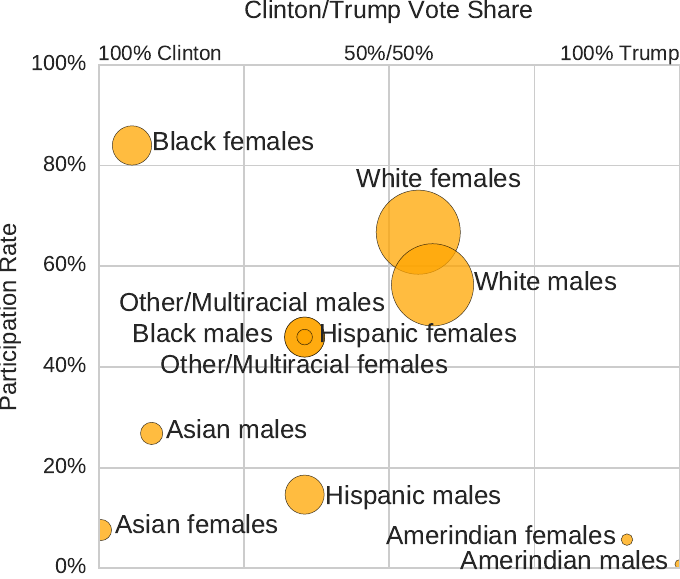}
    \caption{Sex-ethnicity}
\end{subfigure}
\caption{Several categorical features' estimated marginal effects on outcomes.}
\label{fig:categorical}
\end{figure}

In order to explore which parameters were correlated with the outcome,
we used penalized multinomial distribution regression using the same setup as above
restricted to subsets of related categorical parameters, e.g.~there are 6 different categories
for marital status, as shown in Figure \ref{fig:marital}. In each case, we calculated the
crossvalidated deviance for the multinomial logistic likelihood, which is a measure of fit
allowing different models with different numbers of parameters to be compared. Smaller deviance
is better. In Table \ref{table:deviance} we show the top 25 performing features.
% latex table generated in R 3.3.1 by xtable 1.8-2 package
% Thu Nov 10 02:45:56 2016
\begin{table}[ht]
\centering
\begin{tabular}{rlrr}
  \hline
 & feature & deviance & frac.deviance \\ 
  \hline
1 & RAC3P - race coding & 0.04 & 0.86 \\ 
  2 & ethnicity interacted with has degree& 0.04 & 0.74 \\ 
  3 & schooling attainment & 0.04 & 0.72 \\ 
  4 & ANC2P - detailed ancestry & 0.04 & 0.83 \\ 
  5 & OCCP - occupation & 0.04 & 0.75 \\ 
  6 & COW - class of worker & 0.04 & 0.67 \\ 
  7 & ANC1P - detailed ancestry & 0.05 & 0.77 \\ 
  8 & NAICSP - industry code & 0.05 & 0.71 \\ 
  9 & RAC2P - race code & 0.05 & 0.70 \\ 
  10 & age interacted with usual hours worked per week (WKHP) & 0.05 & 0.69 \\ 
  11 & sex interacted with ethnicity& 0.05 & 0.65 \\ 
  12 & MSP - marital status& 0.05 & 0.61 \\ 
  13 & FOD1P - field of degree & 0.05 & 0.61 \\ 
  14 & ethnicity & 0.06 & 0.57 \\ 
  15 & RAC1P - recoded race & 0.06 & 0.54 \\ 
  16 & sex interacted with age & 0.06 & 0.57 \\ 
  17 & has degree interacted with age & 0.06 & 0.55 \\ 
  18 & age interacted with personal income & 0.06 & 0.76 \\ 
  19 & sex interacted with hours worked per week  & 0.06 & 0.62 \\ 
  20 & personal income interacted with hours worked per week & 0.06 & 0.69 \\ 
  21 & personal income& 0.06 & 0.59 \\ 
  22 & RACSOR - single or multiple race & 0.07 & 0.42 \\ 
  23 & has degree interacted with hours worked per week & 0.07 & 0.59 \\ 
  24 & hispanic & 0.07 & 0.56 \\ 
  25 & sex interacted with personal income & 0.07 & 0.57 \\ 
   \hline
\end{tabular}
\caption{Lower deviance values indicate a better fit. Larger fraction of deviance explained values are better. Models are fit to the outcome data as described
in the text using only the categorical predictors or interactions of
categorical predictors listed.}
\label{table:deviance}
\end{table}

\begin{figure}[ht]
    \centering
    \includegraphics[width=.7\textwidth]{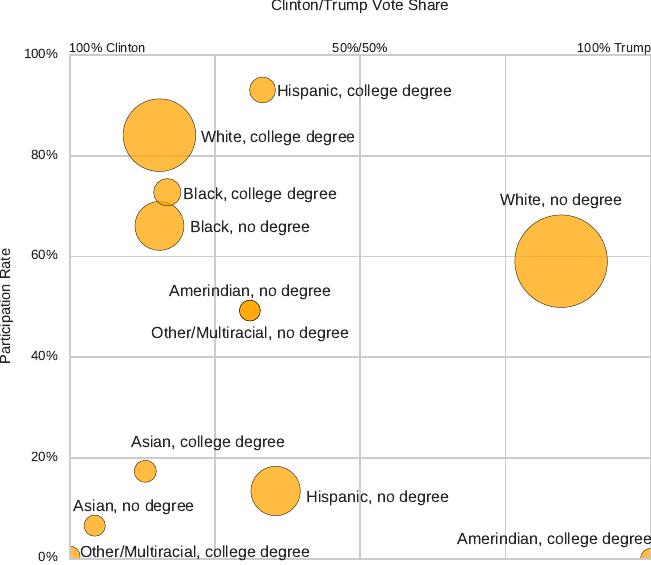}
    \caption{Ethnicity-has degree}
    \label{fig:ethnicity-has-degree}
\end{figure}

We visualize the marginal effects of the most informative categorical features in Figures~\ref{fig:ethnicity-has-degree} and \ref{fig:categorical}, considering not just
their predictive ability for Clinton vs.~Trump but also their predictive ability in terms
of voting for other / not voting, labeling ``Participation Rate'' to mean the fraction of
eligible voters in a category who voted for either Trump or Clinton,
thus combining both nonvoters and those who voted for a third-party candidate.
In Figure \ref{fig:ethnicity-has-degree}, for example, whites with college degrees 
participated highly and mostly supported Clinton. Whites without college degrees did not participate
as highly and mostly supported Trump.
Sizes of each buble represent the approximate relative size of each subgroup.
For race, occupation, and education level, we show only the most-populous subgroups.


\clearpage
\section{Conclusions}
This is a work in progress---we are sharing it before the analysis is complete
in order to get feedback and in the hopes that it can inform the thinking and analysis of others. 
In particular, we intend to make comparisons to 2012 data.
We will also further explore the wealth of variables in the
American Community Survey, add fine-grained spatial maps to
complement the demographic modeling in Section \ref{section:inferring} and
visualize the distributions of real-valued covariates for the most
important features discovered in Section \ref{section:group-based}. Finally, we
will consider Bayesian alternatives to the penalized MLE approach in Section
\ref{section:group-based} so as to, e.g.~infer posterior uncertainty regions
over the location of our estimates in the square plots.

\bibliographystyle{plain}
\bibliography{us2016}

\clearpage
\newpage

\appendix
\section{Model fit}
\label{sec:model-fit}
\begin{figure}[ht]
	\centering
	\includegraphics[width=.6\paperwidth]{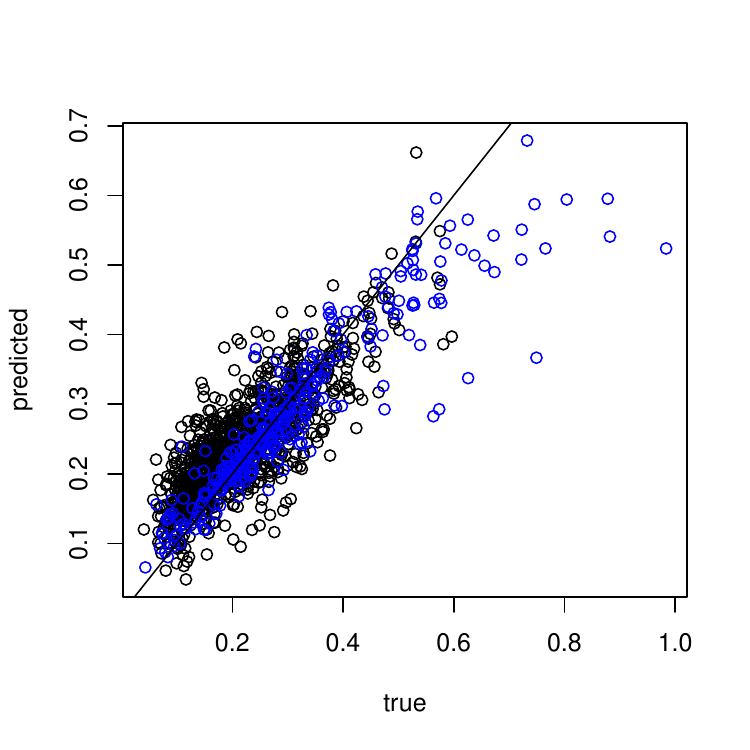}
	\caption{Fit of our model. Black dots are electoral outcomes by local area, while blue dots are exit polling (survey-based) state-level results by age, sex, and education.}
	\label{fig:scatter}
\end{figure}

\clearpage
\section{Supplementary results}
\label{sec:supplement}
% latex table generated in R 3.3.1 by xtable 1.8-2 package
% Thu Nov 10 00:46:16 2016
\begin{table}[ht]
\centering
\begin{tabular}{p{7cm}lrrrrr}
  \hline
 &  Clinton & Trump & Part.& Other / Non-voting \\
  \hline
 Hearing difficulty & 0.37 & 0.63  & 0.47 & 0.53 \\ 
   Vision difficulty & 0.42 & 0.58  & 0.46 & 0.54 \\ 
   Independent living difficulty & 0.50 & 0.50  & 0.28 & 0.72 \\ 
   Veteran service connected disability rating (percentage) $\geq$ 10\% & 0.07 & 0.93  & 0.82 & 0.18 \\ 
   Cognitive difficulty & 0.52 & 0.48  & 0.23 & 0.77 \\ 
   Ability to speak English = well & 0.73 & 0.27  & 0.31 & 0.69 \\ 
   Ability to speak English = not well or not at all  & 0.67 & 0.33  & 0.21 & 0.79 \\ 
   Gave birth to a child within the past 12 months & 0.51 & 0.49  & 0.64 & 0.36 \\ 
   Grandparents living with grandchildren & 0.46 & 0.54  & 0.44 & 0.56 \\ 
  Grandparents responsible for grandchildren & 0.42 & 0.58  & 0.53 & 0.47 \\ 
 Insurance purchased directly from an insurance company  & 0.43 & 0.57  & 0.59 & 0.41 \\ 
  Medicare, for people 65 and older, or people with certain disabilities & 0.44 & 0.56  & 0.46 & 0.54 \\ 
  Medicaid or similar & 0.56 & 0.44  & 0.27 & 0.73 \\ 
  TRICARE or other military health care & 0.26 & 0.74  & 0.72 & 0.28 \\ 
  VA used ever & 0.21 & 0.79  & 0.70 & 0.30 \\ 
  Indian Health Service  & 0.69 & 0.31  & 0.22 & 0.78 \\ 
   \hline
\end{tabular}
\end{table}

% latex table generated in R 3.3.1 by xtable 1.8-2 package
% Thu Nov 10 00:45:33 2016
\begin{table}[ht]
\centering
\begin{tabular}{p{7cm}lrrrrr}
  \hline
 &  Clinton & Trump &  Part & Other / Non-voting \\
  \hline
 Language other than English spoken at home & 0.74 & 0.26 &  0.32 & 0.68 \\ 
  Mobility = lived here one year ago & 0.45 & 0.55 & 0.55 & 0.45 \\ 
  Mobility = moved here from outside US and Puerto Rico & 0.60 & 0.40 & 0.47 & 0.53 \\ 
  Mobility = moved here from inside US or Puerto Rico  & 0.56 & 0.44 & 0.48 & 0.52 \\ 
  Active duty military & 0.45 & 0.55 & 0.56 & 0.44 \\ 
  Not enrolled in school & 0.45 & 0.55 &  0.60 & 0.40 \\ 
  Enrolled in a public school or public college & 0.61 & 0.39 & 0.39 & 0.61 \\ 
  Enrolled in private school, private college, or home school & 0.66 & 0.34 & 0.53 & 0.47 \\ 
   \hline
\end{tabular}
\end{table}

% latex table generated in R 3.3.1 by xtable 1.8-2 package
% Thu Nov 10 00:43:04 2016
\begin{table}[ht]
\centering
\begin{tabular}{p{5cm}lrrrrr}
  \hline
 &  Clinton & Trump & Frac& Part & Other / Non-voting \\
  \hline
 18 $<$= age $<$= 29 \& men & 0.55 & 0.45 & 0.08 & 0.39 & 0.61 \\ 
   18 $<$= age $<$= 29 \& women & 0.71 & 0.29 & 0.08 & 0.37 & 0.63 \\ 
   30 $<$= age $<$= 44 \& men & 0.43 & 0.57 & 0.13 & 0.56 & 0.44 \\ 
   30 $<$= age $<$= 44 \& women & 0.55 & 0.45 & 0.14 & 0.58 & 0.42 \\ 
   45 $<$= age $<$= 64 \& men & 0.48 & 0.52 & 0.16 & 0.49 & 0.51 \\ 
   45 $<$= age $<$= 64 \& women & 0.58 & 0.42 & 0.21 & 0.61 & 0.39 \\ 
   age $>$= 65 \& men & 0.36 & 0.64 & 0.10 & 0.57 & 0.43 \\ 
   age $>$= 65 \& women & 0.49 & 0.51 & 0.10 & 0.48 & 0.52 \\ 
   \hline
\end{tabular}
\end{table}

% latex table generated in R 3.3.1 by xtable 1.8-2 package
% Thu Nov 10 02:05:35 2016
\begin{table}[ht]
\centering
\begin{tabular}{lrrrrr}
  \hline
 & Clinton & Trump & Fraction& Participation & Other \\ 
  \hline
 hispanic  men & 0.66 & 0.34 & 0.04 & 0.39 & 0.61 \\ 
   white  men & 0.39 & 0.61 & 0.36 & 0.57 & 0.43 \\ 
   black  men & 0.77 & 0.23 & 0.05 & 0.44 & 0.56 \\ 
   amerindian  men & 0.50 & 0.50 & 0.00 & 0.28 & 0.72 \\ 
   asian  men & 0.74 & 0.26 & 0.01 & 0.37 & 0.63 \\ 
   hispanic  women & 0.71 & 0.29 & 0.04 & 0.39 & 0.61 \\ 
   white  women & 0.48 & 0.52 & 0.38 & 0.57 & 0.43 \\ 
   black  women & 0.87 & 0.13 & 0.07 & 0.61 & 0.39 \\ 
   amerindian  women & 0.50 & 0.50 & 0.00 & 0.26 & 0.74 \\ 
   asian  women & 0.79 & 0.21 & 0.02 & 0.40 & 0.60 \\ 
   \hline
\end{tabular}
\end{table}

% latex table generated in R 3.3.1 by xtable 1.8-2 package
% Thu Nov 10 00:43:42 2016
\begin{table}[ht]
\centering
\begin{tabular}{lrrrrr}
  \hline
 &  Clinton & Trump & Frac & Part & Other / Non-voting \\
  \hline
 personal income $<$= 50000 \& men & 0.56 & 0.44 & 0.25 & 0.37 & 0.63 \\ 
   personal income $<$= 50000 \& women & 0.63 & 0.37 & 0.36 & 0.40 & 0.60 \\ 
   50000 $<$ personal income $<$= 100000 \& men & 0.40 & 0.60 & 0.15 & 0.67 & 0.33 \\ 
   50000 $<$ personal income $<$= 100000 \& women & 0.53 & 0.47 & 0.13 & 0.84 & 0.16 \\ 
   personal income $>$ 100000 \& men & 0.49 & 0.51 & 0.08 & 0.70 & 0.30 \\ 
   personal income $>$ 100000 \& women & 0.62 & 0.38 & 0.03 & 0.80 & 0.20 \\ 
   \hline
\end{tabular}
\end{table}

\end{document}